\documentclass[prl,cha,onecolumn,superscriptaddress,preprintnumbers,showpacs, amsmath,amssymb]{revtex4-1}
\usepackage{bm}
\usepackage[colorlinks=true,linkcolor=blue,citecolor=blue]{hyperref}
\usepackage{amsmath}
\usepackage{amssymb}
\usepackage{amsthm}
\usepackage{amsfonts}
\usepackage{enumerate}
\usepackage{latexsym}
\usepackage{ifpdf}
\usepackage{graphicx}
\usepackage{makeidx}
\expandafter\ifx\csname package@font\endcsname\relax\else
 \expandafter\expandafter
 \expandafter\usepackage
 \expandafter\expandafter
 \expandafter{\csname package@font\endcsname}
 \fi
\begin{document}

\title{Magnetic interactions at the nanoscale in trilayer titanates LaTiO$_3$/SrTiO$_3$/YTiO$_3$}

\author{Yanwei Cao}
\email{yc003@uark.edu}
\affiliation{Department of Physics, University of Arkansas, Fayetteville, AR 72701, USA}
\author{Zhenzhong Yang}
\affiliation{Beijing National Laboratory for Condensed Matter Physics and Institute of Physics, Chinese Academy of Sciences, Beijing 100190, P. R. China}
\author{M. Kareev}
\affiliation{Department of Physics, University of Arkansas, Fayetteville, AR 72701, USA}
\author{Xiaoran Liu}
\author{D. Meyers}
\affiliation{Department of Physics, University of Arkansas, Fayetteville, AR 72701, USA}
\author{S. Middey}
\affiliation{Department of Physics, University of Arkansas, Fayetteville, AR 72701, USA}
\author{D.  Choudhury}
\affiliation{Department of Physics, University of Arkansas, Fayetteville, AR 72701, USA}
\affiliation{Department of Physics, Indian Institute of Technology, Kharagpur 721302, India}
\author{P. Shafer}
\affiliation{Advanced Light Source, Lawrence Berkeley National Laboratory, Berkeley, California 94720, USA}
\author{Jiandong Guo}
\affiliation{Beijing National Laboratory for Condensed Matter Physics and Institute of Physics, Chinese Academy of Sciences, Beijing 100190, P. R. China}
\affiliation{Collaborative Innovation Center of Quantum Matter, Beijing 100190, P. R. China}
\author{J. W. Freeland}
\affiliation{Advanced Photon Source, Argonne National Laboratory, Argonne, Illinois 60439, USA}
\author{E. Arenholz}
\affiliation{Advanced Light Source, Lawrence Berkeley National Laboratory, Berkeley, California 94720, USA}
\author{Lin Gu}
\affiliation{Beijing National Laboratory for Condensed Matter Physics and Institute of Physics, Chinese Academy of Sciences, Beijing 100190, P. R. China}
\affiliation{Collaborative Innovation Center of Quantum Matter, Beijing 100190, P. R. China}
\author{J. Chakhalian}
\affiliation{Department of Physics, University of Arkansas, Fayetteville, AR 72701, USA}

\date{\today}

\begin{abstract} 

We report on the phase diagram of competing magnetic interactions at nanoscale in engineered ultra-thin trilayer heterostructures of LaTiO$_{3}$/SrTiO$_{3}$/YTiO$_{3}$,  in which the interfacial inversion symmetry is explicitly broken. Combined atomic layer resolved scanning transmission electron microscopy with electron energy loss spectroscopy and electrical transport have confirmed the formation of a spatially separated two-dimensional electron liquid and high density two-dimensional localized magnetic moments at the LaTiO$_3$/SrTiO$_3$ and SrTiO$_3$/YTiO$_3$ interfaces, respectively. Resonant soft X-ray linear dichroism spectroscopy has demonstrated the presence of orbital polarization of the conductive LaTiO$_3$/SrTiO$_3$ and localized SrTiO$_3$/YTiO$_3$ electrons. Our results provide a route with prospects for exploring new magnetic interfaces, designing tunable two-dimensional $d$-electron Kondo lattice, and potential spin Hall applications.

\end{abstract}

\maketitle

Magnetic interactions between the localized spins and conduction electrons are a fundamental component  in  the plenty of intriguing quantum many-body phenomena \cite{PTP-1964-Kondo,RMP-1997-Tsu,2007-Coleman,Nmat-2012-Coleman,Nmat-2014-Kim,PRL-2009-Guo,PRB-2008-Gorini,Nphy-2013-Gabay,PRB-2014-Ruh,Hewson-1993-Kondo,ZPB-1991-Fazekas,RMP-2006-Jun,science-2010-Yu,science-2013-Chang,MRS-2013-Coey,Nmat-2007-Brinkman1}, a remarkable manifestation of which is the Kondo effect in heavy fermion systems \cite{RMP-1997-Tsu,2007-Coleman,Nmat-2012-Coleman,ZPB-1991-Fazekas,Hewson-1993-Kondo}. Phenomenologically, in  systems with localized spins coupled to conduction electrons,  the Kondo interaction\cite{PTP-1964-Kondo,Hewson-1993-Kondo} competes with the magnetic Rudderman-Kittel-Katsuya-Yosida (RKKY) interaction \cite{RMP-1997-Tsu} leading to the Doniach phase diagram \cite{PB-1977-Doniach,PRL-1999-Sullow} and Kondo lattice models \cite{RMP-1997-Tsu,2007-Coleman,Nmat-2012-Coleman,ZPB-1991-Fazekas}. In real transition metal compounds, however, the ground state strongly depends on the exchange interaction ($J$), electronic density ratio (n$_m$/n$_c$) of the localized magnetic moments (n$_m$) to conduction electrons (n$_c$), and the orbital character of magnetically active electrons 
\cite{RMP-1997-Tsu,RMP-2006-Jun,ZPB-1991-Fazekas,Nphy-2013-Gabay,PRB-2014-Ruh}.  In the strong-coupling regime with large $|J|$, the Kondo interaction prevails with the formation of a Kondo singlet state \cite{PTP-1964-Kondo},  whereas on the weak-coupling side (small $|J|$) depending on the value of n$_m$/n$_c$ \cite{RMP-2006-Jun,ZPB-1991-Fazekas} the RKKY interaction may give rise to either  a ferromagnetic (FM) or antiferromagnetic (AFM)\ order between the localized spins. In particular, in the limiting case of n$_m$/n$_c\gg1$, the localized spins  tend to form ferromagnetic order by polarizing the conduction electrons via the Zener kinetic exchange mechanism \cite{PR-1951-Zener,RMP-2006-Jun,PRL-2012-Mich}.

In correlated \textit{d}-electron heterointerfaces, besides the density ratio n$_m$/n$_c$, the  dimensionality and orbital polarization of the magnetic interactions are all vital components for the formation of a  ground state as exemplified by the emergent FM at the manganate-cuprates  \cite{NP-2006-JC,Science-2007-JC}and manganite-ruthenate interfaces \cite{PRB-2010-Freeland,PRB-2011-Yor,PRL-2012-He,PRL-2015-Grutter}. Moreover, it has been revealed  that due to the significantly enhanced quantum fluctuations in reduced  dimensions, under the scenario of Kondo state destruction \cite{Nmat-2012-Coleman,Science-2010-Shishido,Nphy-2011-Mizukami} the electron and spin degrees of freedom start playing a major role to tune the heavy-fermion metal into a magnetic metal phase by crossing the quantum critical point. On the other hand, compared to $f$-electron heavy fermion compounds, complex oxides with partially filled \textit{d}-shells are believed to  be the most promising candidates for high-temperature quantum materials  \cite{Nmat-2012-Coleman}, since the \textit{d}-electron \textit{orbital} configuration  intimately defines their magnetic ground states \cite{Khomskii-2014,Science-2000-Tokura}. To emphasize this, consider the splitting of the Ti t$_{2g}$ band between $d_{xy}$- and $d_{xz}$/$d_{yz}$-subbands,  which is  a prime cause for the interesting emergent phenomena in the SrTiO$_3$-based heterostructures \cite{Nmat-2012-Hwang,Nmat-2012-JC,ARCMP-2011-Zubko,Science-2010-Man,MRS-2013-Gra,ARMR-2014-Stemmer,MRS-2008-Mannhart,NJP-2014-Bja} including the coexistence of superconductivity and ferromagnetism \cite{PRL-2012-Mich,NC-2012-Kal,NP-2011-Bert,NP-2011-Li,NP-2013-Ban}, the appearance of interfacial ferromagnetism \cite{Nphy-2013-Gabay,PRB-2014-Ruh,APL-2011-Moe,Nmat-2007-Brinkman1,MRS-2013-Coey,NC-2012-Kal,NP-2013-Ban,RMP-2014-JC,NC-2010-GB,NM-2011-SV,NP-2011-Bert,NP-2011-Li}, and the formation of  one-dimensional bands \cite{MRS-2013-Coey}. Since the density ratio n$_{m}$/n$_{c}$, spatial confinement, and $d$-electron orbital character are all important for activating novel or latent quantum states, it raises an important question:  what is an experimental phase diagram for the emerging magnetic interactions at nanoscale?

\begin{figure*}[t]
\includegraphics[width=0.9\textwidth]{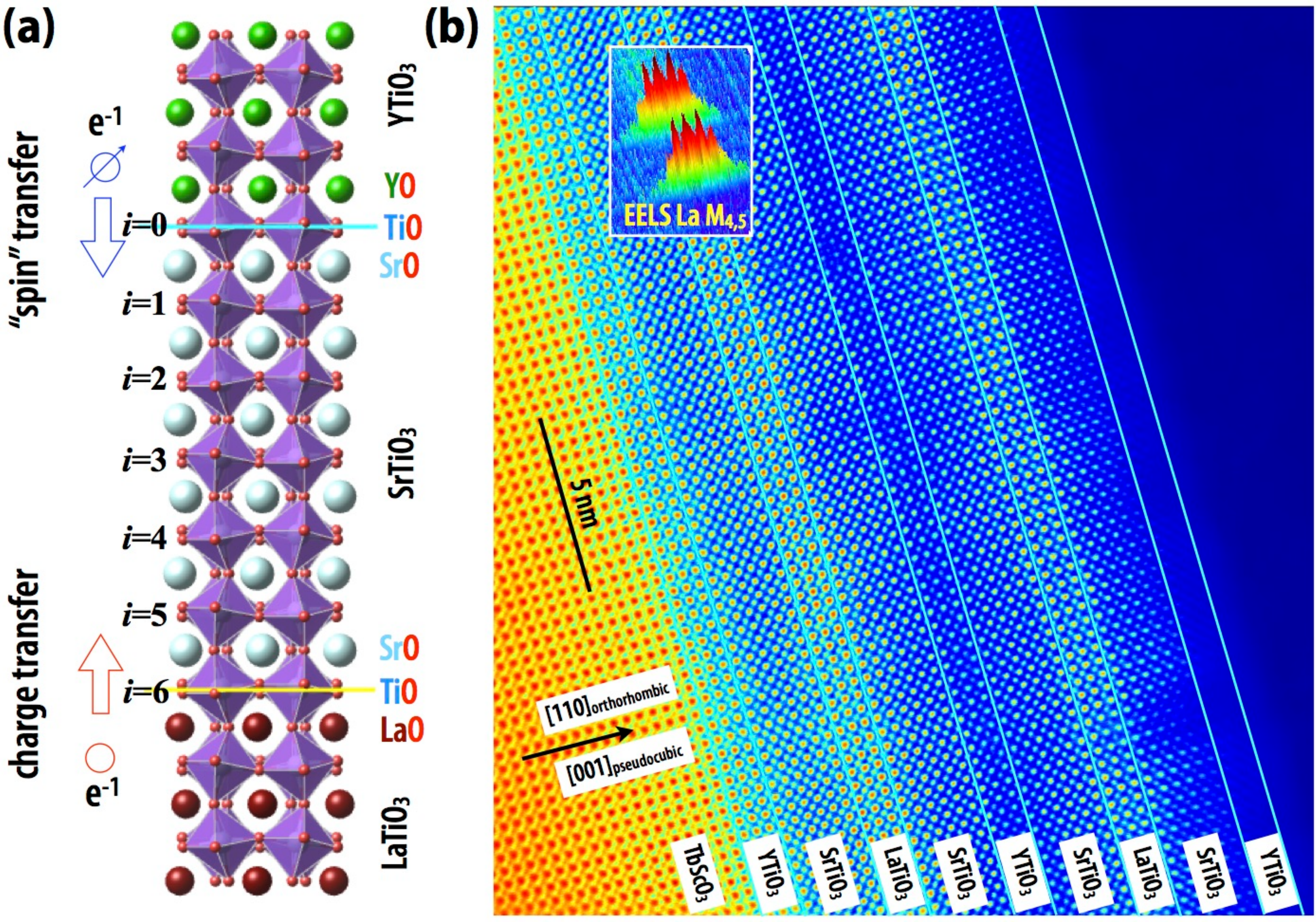}
\caption{\label{} (Color online) (a) Schematic view of 3LTO/6STO/3YTO with two active heterointerfaces: LTO/STO and YTO/STO. Here $i$ denotes the $i^{\textrm{th}}$ TiO$_2$ atomic plane of the STO layer with $n$ unit cells. (b) HAADF-STEM image of 3LTO/6STO/3YTO. The inset is EELS spectra at La M$_{4,5}$-edge, the three main peaks of which demonstrate the atomic sharpness of interfaces.}
\end{figure*}

To address this challenge, we synthesized a class of asymmetric trilayer superlattices (SLs) [$m$ u.c. LaTiO$_{3}$/$n$ u.c. SrTiO$_{3}$/$t$ u.c. YTiO$_{3}$] (thereafter $m$LTO/$n$STO/$t$YTO, u.c.~=~unit cells; see Fig.~1). In this system, the interfacial charge-transfer was utilized to create a two-dimensional (2D) Fermi liquid interface (LTO/STO) and a spatially separated interface with localized magnetic moments (YTO/STO). This charge transfer across the layers was evidenced by combined scanning transmission electron microscopy (STEM) with electron energy loss spectroscopy (EELS) and electrical transport. Resonant soft X-ray linear dichroism (XLD) measurements were used to probe the orbital polarization of n$_{m}$ and n$_{c}$ electrons. The results have allowed to map  out  the phase diagram of competing and altered magnetic interactions at nanoscale by changing the distance between the two  electronically  active interfaces via the STO layer thickness $n$.

Trilayer SLs [$3$LTO/$n$STO/$3$YTO] $\times$ 4 ($n$~=~2, 3, 6) and reference samples $m$LTO/$n$STO ($m$~=~3, 20 and $n$~=~2, 3, 6) and 3YTO/$n$STO ($n$~=~2 and 6) were epitaxially synthesized on TbScO$_3$ (110) substrates by pulsed laser deposition (PLD) with a layer-by-layer mode (see details for the growth of single layer LTO, STO, and YTO films in our previous reports) \cite{APL-2013-Misha,APL-2015-YC}. A JEM-ARM200F STEM, operated at 200 kV and equipped with double aberration-correctors for both probe-forming and imaging lenses, was used to perform high-angle annular-dark-field (HAADF) imaging and EELS spectroscopy. The sheet-resistances of the films were measured in van-der-Pauw geometry by Physical Properties Measurement System (PPMS, Quantum Design). XLD at Ti L$_{2,3}$-edge with total electron yield (TEY) mode was carried out at beamline 4.0.2 (using the vector magnet) of the Advanced Light Source (ALS, Lawrence Berkeley National Laboratory).

\begin{figure*}[t]
\includegraphics[width=0.9\textwidth]{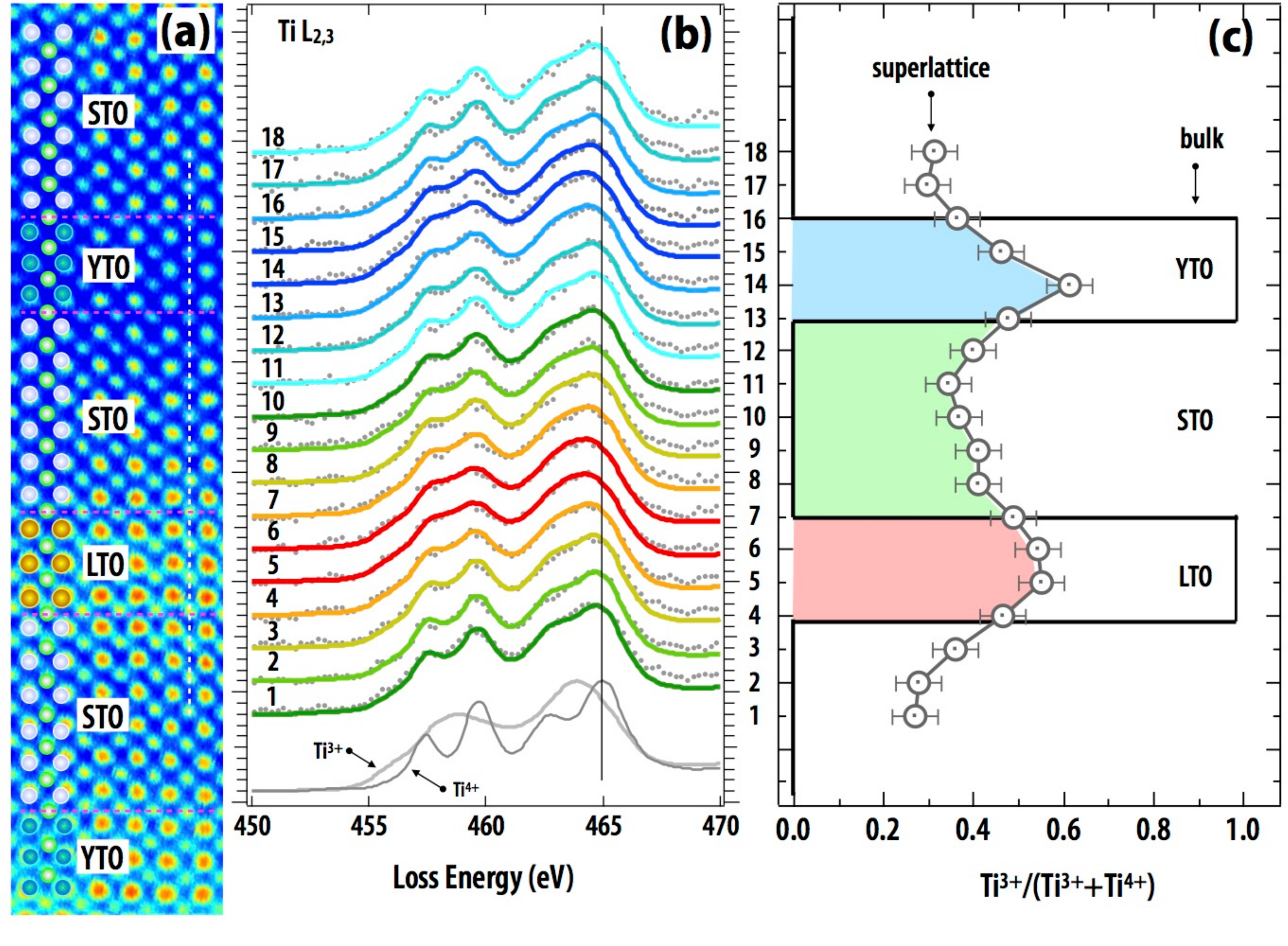}
\caption{\label{} (Color online). (a) HAADF image of 3LTO/6STO/3YTO. The atomic positions of the elements La (large yellow dots), Sr (large white dots), Y (large blue dots), Ti (small green dots) are labeled schematically. (b) Layer-resolved EELS spectra (scanning along the dashed white lines in (a)) with the left side of the spectra aligned to the HAADF image in (a). Reference spectra for Ti$^{4+}$ and Ti$^{3+}$ (acquired on bulk-like SrTiO$_3$ and LaTiO$_3$ films, respectively) at the bottom were adapted from Ref. 50. The colored solid lines are the fitting curves of layer-resolved EELS spectra to a linear combination of Ti$^{4+}$ and Ti$^{3+}$ spectra, whereas the experimental data are marked by small gray dots. (c) Spatial decay of the Ti$^{3+}$‡ signal across the two heterointerfaces. The Ti$^{3+}$‡ spectra weight is estimated from the fitting parameters of solid curves in (b).}
\end{figure*}

As shown in Fig.~1(a), in the trilayer heterostructure LTO/STO/YTO there are two inequivalent interfaces composed of two rare- and one alkaline-earth titanate compounds (LTO, YTO and STO) with  a rich phase diagram \cite{NJP-2004-Mas}. In the bulk form, the Mott-insulator LaTi$^{3+}$O$_{3}$ ($\sim$ 0.2~eV gap, Ti 3\textit{d}$^{1}$) undergoes a G-type AFM phase transition below 146~K, while the Mott-insulator YTi$^{3+}$O$_3$ ($\sim$ 1.2~eV gap, Ti 3\textit{d}$^{1}$) is FM below 30~K \cite{APL-2015-YC,NJP-2004-Mas}, and, finally,  SrTi$^{4+}$O$_3$ remains a band-insulator ($\sim$ 3.2~eV gap, Ti 3\textit{d}$^{0}$) over the whole temperature range. The 2D conduction electrons at the LTO/STO interface \cite{Nature-2002-Ohtomo,APL-2013-Misha}, resulting from the charge-transfer from the LTO into STO layers (red arrow in Fig.~1(a)), serve as the Fermi sea whereas the \textquotedblleft spin\textquotedblright-transfer from YTO into STO (blue arrow in Fig.~1(a)) at the other interface (YTO/STO) produces the 2D localized magnetic moments and induces spin polarization in the interface \cite{Science-2011-Jang2}. To assure that the magnetic interactions can be investigated at the unit cell scale, the interface roughness was investigated by (HAADF) STEM imaging with atomic resolution which gives the structural projection of the SLs. Taking 3LTO/6STO/3YTO as a representative example, the geometries and sequences of the trilayer layers LTO, STO and YTO in these SLs are clearly seen with atomically sharp interfaces as displayed in Fig.~1(b) and Fig. S1.

To check for the presence of the interfacial charge-transfer across the two LTO/STO and STO/YTO interfaces, the layer-resolved electronic structure of 3LTO/6STO/3YTO was investigated by atomic scale STEM-EELS line scanning  across  the interfaces. As seen in Fig.~2, by scanning atomic layer-resolved Ti L$_{2,3}$-edge spectra across LTO/STO and STO/YTO interfaces (along the white-dashed line in Fig.~2(a)) with high energy (0.4~eV) and spatial (0.8~\AA) resolutions, the evolution of the Ti electronic structure through the interfaces was clearly observed (Fig.~2(b)). 
Additionally, atomic layer-dependent lineshape and peak positions of the EELS spectra carry important information regarding the interfacial charge-transfer. 
Since each curve of the Ti L$_{2,3}$-edge spectra is  a convolution of both Ti$^{3+}$ and Ti$^{4+}$ spectra, the spectral weight of Ti$^{3+}$ in each spectral line of Ti$^{3+}$/(Ti$^{3+}$+Ti$^{4+}$) can track and quantify the process of interfacial charge-transfer. 
The direct inspection of the EELS\ data revealed that in addition to the  the previously reported charge-transfer from LTO into STO \cite{Nmat-2012-Hwang,Nature-2002-Ohtomo} there is an unexpectedly large charge-transfer from YTO into STO (see Fig.~2(c)) which  leads to  a localized electron layer formation at the YTO/STO interface \cite{APL-2013-Misha,Science-2011-Jang2}.

\begin{figure*}[t]
\includegraphics[width=0.9\textwidth]{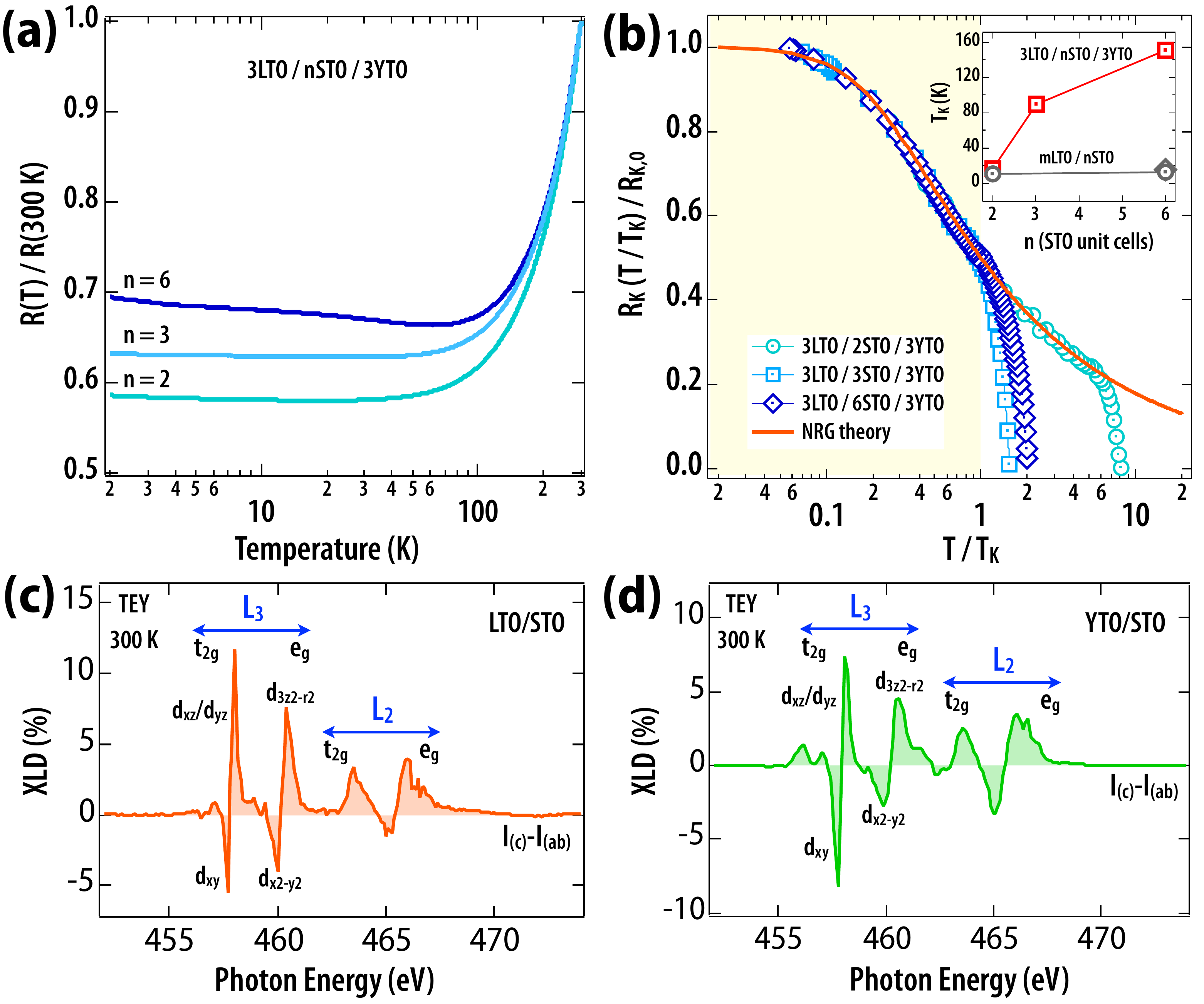}
\caption{\label{} (Color online). (a) Normalized sheet resistances $R(T)/R(300K)$ of 3LTO/$n$STO/3YTO. (b) Experimentally and theoretically [red solid line, numerical renormalization group (NRG)] scaled Kondo resistances [$R_{K}(T/T_{K})/R_{K,0}$] (see Eq.~S4 for details). Inset, extracted $n$-dependent Kondo temperature $T_{K}$ by fitting the experimental data of $m$LTO/$n$STO and 3LTO/$n$STO/3YTO (see Fig.~S3). (c) and (d) XLD [$I(c)-I(ab)$] of 3LTO/10STO and 3YTO/2STO interfaces with surface-sensitive TEY mode, where $I(c)$ [$E\parallel c$, $E$ is the polarization vector of the photon] is for out-of-plane and $I(ab)$ [$E\parallel ab$] is for in-plane detecting. It is noted most contribution of the signal is from top few layers (STO).}
\end{figure*}

\begin{figure*}[t]
\includegraphics[width=0.9\textwidth]{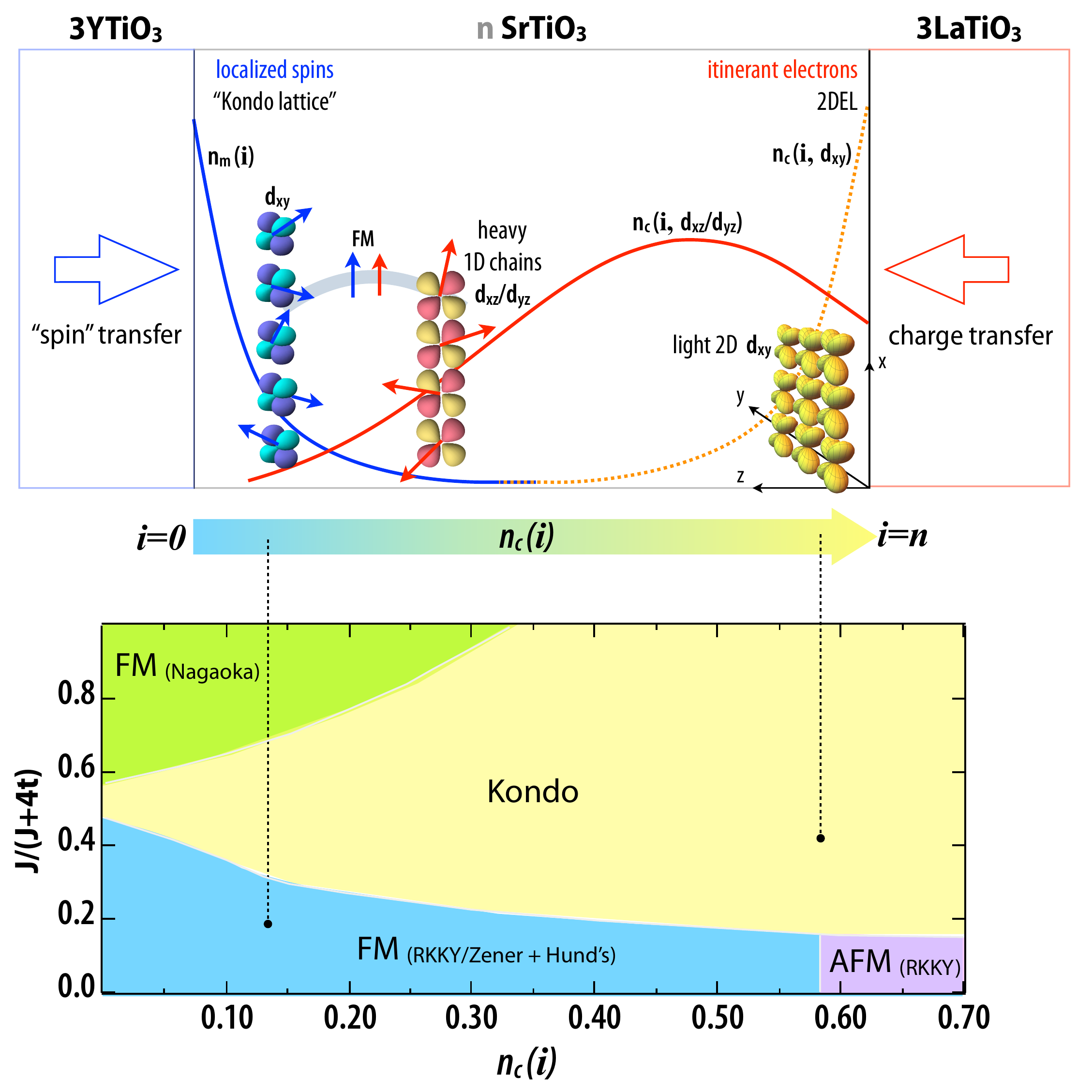}
\caption{\label{} (Color online) Top panel, a sketch of the TiO$_2$ plane ($i$)-dependent electronic density [n$_c$($i$) for itinerant electrons and n$_m$($i$) for localized spins, $i$~$\leq$~$n$ (see Fig.~1(a))] and orbital occupancy with a charge decay away from the interfaces (see also Fig.~2(c) and Fig.~S5\cite{sup}). Bottom panel, schematic phase diagram adapted from the theory \cite{RMP-2006-Jun,ZPB-1991-Fazekas}. Near the LTO/STO interface the Kondo effect is dominant whereas ferromagnetic exchange is favored near the YTO/STO interface forming 2D $d$-electron Kondo lattice-like structure. Note, the phase diagram is effective for both periodic and random localized magnetic moments \cite{RMP-2006-Jun,ZPB-1991-Fazekas}. Here, $t$ is the electron hopping energy. }
\end{figure*}

Next we investigated the properties of interfacial electrons (arising from the interfacial charge-transfer) by measuring the temperature-dependent electrical transport. As seen in Fig.~3(a) and Fig.~S2\cite{sup}, all  three samples 3LTO/$n$STO/3YTO ($n$~=~2, 3, 6) show characteristic metallic behavior with a weak upturn at lower temperature. To investigate the conducting properties of the two interfaces (LTO/STO and STO/YTO) in the trilayer SLs and rule out  possible contributions from defects and oxygen vacancies,  \textit{bilayer} YTO/STO and LTO/STO samples were synthesized and their transport properties  were used as references (see Fig.~S2); In sharp contrast to highly insulating YTO/STO \cite {APL-2013-Misha,Science-2011-Jang2}, the sheet resistances of all the LTO/STO samples (Fig.~S2(a)\cite{sup}) show a 2D electron liquid (2DEL) behavior \cite{PRL-2013-Chang,Nature-2002-Ohtomo,NC-2010-Bis,Science-2011-Jang2,NM-2013-Bis}.

With the creation of 2D conduction electrons at LTO/STO interface, the other ingredient,  that needs to be considered towards the realization of controlled magnetic interactions between two separately active heterointerfaces, is the formation of localized spins at the STO/YTO junction. 
As highlighted by the normalized sheet resistances in Fig.~S2(d)\cite{sup} (cyan shadow area, marked as $T_\textrm{m}$) and Fig.~3(a), the other dominant feature in transport  is the pronounced upturn  of the sheet resistances at lower temperature.
Previous work on titanate heterojunctions has attributed such an upturn to the Kondo effect \cite{Nmat-2007-Brinkman1,PRB-2014-De1,AMI-2014-Lin1,PRB-2014-Das1,PRB-2014-Das2}, after carefully ruling out the contributions from weak-localization \cite{Gan-2005} and electron-electron interactions \cite{PRB-2013-Chiu}. 
One of the key features of the Kondo effect that immediately differentiates it from weak-localization and electron-electron interactions is the \textit{universal scaling behavior} [see Fig. S3-S4 and Eq.~(S1)-(S4)\cite{sup}].  As shown in Fig.~3(b), all the \textit{trilayer} SLs obey the Kondo scaling behavior which further affirms the dominant contribution of Kondo screening to the observed upturn feature.
Based on the observation that YTO/STO interface shows massive charge-transfer and yet is highly insulating, the observed Kondo  behavior lends strong evidence for the formation of interfacial localized magnetic moments located on the STO side proximal to the YTO/STO interface of LTO/STO/YTO. 
The combined STEM/EELS and electrical transport data thus established that a 2D conduction electron layer is formed at the  LTO/STO interface  whereas a high-density 2D localized magnetic layer is formed in the vicinity of the STO/YTO\ interface (see Fig. 2 and Fig. S5).

To elucidate  the link between magnetism and the $d$-orbital occupancy of Ti ions in the STO\ layer, we carried out XLD measurements to probe the orbital character of both itinerant (n$_c$) and localized (n$_m$) electrons, as shown in Fig.~3(c)-(d) and Fig.~S6\cite{sup}.
Notice, to date the study of orbital polarization and subband splitting of interfacial Ti 3$d$ band by XLD were mostly carried out on LaAlO$_3$/SrTiO$_3$ (LAO/STO) \cite{PRL-2009-Sal,AM-2013-Sal,PRL-2013-Sal,PRL-2014-Pes,Nmat-2013-Lee} and very little is known experimentally about  orbital  physics at the LTO/STO and, in particular, the YTO/STO interface. 
As shown in Fig.~3(c)-(d) and Fig.~S6\cite{sup}, the orbital occupancy at both LTO/STO and YTO/STO heterointerfaces exhibits a similar configuration, where the $d_{xy}$ subband is the lowest occupied state compared to the energy position of the $d_{xz}$/$d_{yz}$ subband.
This  orbital configuration is in a good agreement  with the reported  orbital physics at the LAO/STO interface \cite{PRL-2009-Sal,AM-2013-Sal,PRL-2013-Sal,PRL-2014-Pes,Nmat-2013-Lee}. 

To fully understand the magnetic interactions in LTO/STO/YTO, which strongly depend on $J$, n$_m$/n$_c$ and  orbital polarization, the STO\ layer thickness ($n$)-dependent electronic density distribution and \textit{d}-orbital occupancy were  investigated.  As shown in Fig.~S3-S5\cite{sup}, from the combined  EELS and  transport data for both LTO/STO and YTO/STO interfaces, the presence of  high-density of 2D conduction electrons  (n$_c$$\sim$$10^{14}$-$10^{15}$ cm$^{-2}$/interface) and localized moments (n$_m$$\sim$$10^{14}$-$10^{15}$ cm$^{-2}$/interface)\ was deduced. \textcolor{red}{} To demonstrate the dependence of magnetic interactions on n$_m$/n$_c$ and orbital polarization we define the  $i$-dependent carrier densities n$_c$($i$) and n$_m$($i$), where $i$ denotes the $i^{\textrm{th}}$ TiO$_2$ atomic plane inside the STO layer counted from the YTO/STO interface (Fig. 1(a)). As seen in Fig.~2(c) and Fig.~S5, based on the evolution of Ti$^{3+}$/(Ti$^{3+}$+Ti$^{4+}$) ratio, both n$_c$($i$) and n$_m$($i$) show rapid decay behavior into  the STO\  layer.

Next we turn our attention to the orbital character of n$_c$($i$) and n$_m$($i$). Previous extensive\ work on  STO-based heterostructures with 2DEL revealed the presence of two kinds of mobile  carriers with distinct orbital  dispersion, namely  light $d_{xy}$ and heavy $d_{xz}$/$d_{yz}$ electrons\cite{PRL-2013-Chang,PRL-2005-Popovic,PRB-2013-You,Nphy-2013-Gabay}. 
For the LTO/STO interface recent angle-resolved photoemission spectroscopy (ARPES) results \cite{PRL-2013-Chang} confirmed that light $d_{xy}$ conduction electrons with large carrier density n$_c$($i$, $d_{xy}$) are bound to the LTO/STO interface while  away and deeper into  the STO\  layer the mobile carriers are heavy electrons with $d_{xz}$/$d_{yz}$  dispersion (see Fig.~4  top panel).  On the other hand, for YTO/STO no experimental ARPES\ data are available.  First principle\ calculations predict that the YTO/STO interface is ferromagnetic and insulating \cite{Science-2011-Jang2}. 
A direct comparison of  the XLD lineshapes taken on Ti L$_{2,3}$-edge for YTO/STO, LTO/STO, and LAO/STO \cite{PRL-2009-Sal,AM-2013-Sal,PRL-2013-Sal,PRL-2014-Pes,Nmat-2013-Lee}  lends  strong support to  the notion that the $d_{xy}$ band is indeed the lowest occupied state of magnetic \textit{d}-electrons at the YTO/STO interface.

With the observation of  n$_m$($i$)/n$_c$($i$)  and the presence of orbital polarization, it is interesting to speculate how the magnetic interactions  are modulated by the STO\ thickness ($n$). 
We start by considering the case of a thick \textit{n} = 6 STO layer. With a thicker STO layer, near the LTO/STO interface n$_{c}$  ($d_{xy}$)~$\gg$~n$_m$  resulting in the formation of a Kondo singlet state, as illustrated in  Fig.~4 (bottom). On the other hand, a low concentration of heavy electrons with $d_{xz}$/$d_{yz}$ character disperses away from the LTO/STO interface and  appears near the magnetic STO/YTO interface; upon reaching the STO/YTO interface the heavy electrons   interact with  the localized magnetic  moments with $d_{xy}$ character. The orbital-dependent ferromagnetic interactions then can proceed through two possible mechanisms: 
(i) based on the Hund's rule, the interaction between the $d_{xy}$ and $d_{xz}$/$d_{yz}$ electrons results in FM ground state \cite{Nphy-2013-Gabay,PRB-2014-Ruh} and  (ii) the Zener kinetic exchange, which may win the competition with the Kondo and RKKY interactions, again leads to the formation of a localized ferromagnetic ground state with \textit{spin-polarized conduction electrons} \cite{RMP-2006-Jun,ZPB-1991-Fazekas}. 
Based on this  consideration,  both the Hund's rule and Zener kinetic exchange  favor the formation of localized ferromagnetism and spin-polarized conduction carriers.
At the  other limit when the STO layer is ultra-thin (e.g., $n$ = 2),  n$_{c}$~$\sim$~n$_m$ the  conduction carriers  lose their distinct orbital character resulting in the mixed orbital state $d_{xz}$/$d_{yz}/d_{xy}$. In this case the ground state is the result of a direct competition between the Kondo screening, RKKY coupling, and Hund's  energy. Based on this picture,  the control of STO thickness $n$ enables the remarkable ability to modulate the critical  ratio of n$_{c}$/n$_m$ and orbital polarization to exert  definitive control over the magnetic interactions.

In conclusion, we investigated the magnetic coupling between the 2D conduction electrons formed at the LTO/STO interface and the 2D localized magnetic moments at the STO/YTO interface. Due to the STO layer thickness dependent electronic density and orbital polarization of the both conduction electrons and localized spins, the competing Kondo singlet state and ferromagnetism are both present at the different TiO$_2$ planes inside the STO layer of LTO/STO/YTO.  Strength of the magnetic interactions  and the resulting magnetic ground state can be very effectively modulated by the thickness of  STO layer with the nanoscale precision. Our findings provides an emerging magnetic phase diagram which  should  open prospects for  exploring new magnetic  interfaces \cite{Nphy-2013-Gabay,PRB-2014-Ruh}, designing tunable 2D Kondo lattices \cite{Nmat-2012-Coleman,ZPB-1991-Fazekas}, and potential spin Hall applications with complex oxides \cite{PRL-2009-Guo,PRB-2008-Gorini, PRB-2008-Tan}.

The authors deeply acknowledge numerous insightful theory discussions with Andrew Millis, Se Young Park, Gregory Fiete, and Daniel Khomskii. J. C. and D. M. were supported by the Gordon and Betty Moore Foundation EPiQS Initiative through Grant No. GBMF4534. L. Gu and J. Guo acknowledge support from National Basic Research Program of China (Grants No. 2012CB921702 and No. 2014CB921002). X. L. was supported by the Department of Energy Grant No. DE- SC0012375 for his synchrotron work. Y. C., S. M., and M.K. were supported by the DOD-ARO under Grant No. 0402-17291. L. Gu and J. Guo acknowledge the support from the National Natural Science Foundation of China (51522212, 51421002, 11225422) and the Strategic Priority Research of the Chinese Academy of Sciences (Grant No. XDB07030200). The Advanced Light Source is supported by the Director, Office of Science, Office of Basic Energy Sciences, of the U.S. Department of Energy under Contract No. DE-AC02- 05CH11231. This research used resources of the Advanced Photon Source, a U.S. Department of Energy (DOE) Office of Science User Facility operated for the DOE Office of Science by Argonne National Laboratory under Contract No. DE-AC02-06CH11357.

\clearpage

\large{\textbf{Supplementary Materials Online\\
Magnetic interactions at the nanoscale in trilayer titanates LaTiO$_3$/SrTiO$_3$/YTiO$_3$, by Cao \textit{et al.}}}

\subsection{The Kondo effect in 2DEL system}

The upturn behavior of the sheet resistances (with saturation at ultra-low temperature) in high-quality LTO/STO and LAO/STO heterostructures has been assigned to the Kondo effect \cite{Nmat-2007-Brinkman,PRB-2014-De,PRB-2014-DasJ1,PRB-2014-DasJ2,AMI-2013-Lin} after carefully ruling out weak localization and electron-electron interaction. In our work, the size of T$_{m}$ ($\sim$~15 K, temperature with the minimum resistance) for LTO/STO interfaces shows good agreement with the value reported by Das \textit{et al.} \cite{PRB-2014-DasJ1,PRB-2014-DasJ2}. Generally, the Kondo temperature can be expressed by\cite{Hewson-1993}
\begin{align}
k_{B}T_{K} \sim De^{-1/[2|J|\rho(E_{F})]},\\
|J|=|V|^{2}(\frac{1}{|\epsilon_{d}|}+\frac{1}{|U+\epsilon_{d}|}),
\end{align}
where $k_{B}$ is Boltzmann's constant, $D$ is the electronic bandwidth, $J$ for the antiferromagnetic exchange between the transferred-localized spins and conduction electrons in 2DEL, $\rho(E_{F})$ is the density of states of the conduction electrons near the Fermi energy, $|V|^2$ is the hybridization parameter, U is the Coulomb repulsive energy, and $\epsilon_{d}$ is the binding energy of the localized spins. Under the numerical renormalization group (NRG) theory\cite{JPCM-1994-Costi}, it was illustrated that the electrical resistance induced by the Kondo effect (Kondo resistance) can be expressed by a universal function with only a single parameter T/T$_{K}$. The fitting quality of the Kondo resistance is usually taken as a key proof to verify the Kondo scenario\cite{Science-2000-Wiel,Nphys-2011-Chen}. With the NRG theory, the function of the Kondo resistance can be obtained from experimental resistance data by\cite{AMI-2013-Lin,PRL-1998-GG,PRL-2011-Lee}
\begin{align}
R_{s}(T)=&R_{0}+aT^{b}+R_{K}(T/T_{K}),\\
R_{K}(T/T_{K})=&R_{K,0}(\frac{1}{1+(2^{1/S}-1)(T/T_{K})^2})^S,
\end{align}
where $T_{K}$ is the Kondo temperature, $R_{0}$ is the residual resistance, the term $aT^{b}$ is the general expression of transport mainly arising from electron-electron and electron-phonon contribution, $R_{K}(T/T_{K})$ is an empirical function for the universal Kondo resistance as a function of $T/T_K$, $R_{K,0}$ is the Kondo sheet resistance at zero temperature, and the parameter $S$ is fixed at 0.225 for spin $s=\frac{1}{2}$ systems.
With Eq. (S3) and (S4), Figure~S3 shows good agreements between the experimental data and the fitting curves with Kondo scenario for all the bilayer and trilayer samples. Furthermore, as seen in the Fig.~3b of the main text, the Kondo temperatures (T$_{K}$) and the normalized Kondo resistance ($R_{K}(T/T_{K})$/$R_{K,0}$) can be extracted with the fitting parameters. On the other hand, with the limitation $T$ $\sim$ 0, the density of the localized spins leading to Kondo effect can be roughly estimated by\cite{AMI-2013-Lin}
\begin{align}
R_{K,0} \approx \frac{hn_{m,K}}{\pi e^2n_{s}},
\end{align}
where $n_{s}$ and $n_{m,K}$ are the sheet concentrations of conduction electrons and localized moments, respectively.
Therefore, based on the density of the conduction carriers extracted from the Hall measurements, the Kondo sheet resistances near zero temperature and the density of the localized spins can be roughly estimated with Eq. (S5), as shown in Fig. ~S3. Comparing with the bilayer SLs $m$LTO/$n$STO, the increased Kondo magnetic screening centers in trilayer SLs 3LTO/$n$STO/3YTO is consistent with the appearance of the \textquotedblleft spin\textquotedblright~transfer from YTO to STO.

It is important  to note,  the extracted values  of $T_{K}$ (e. g., $\sim$~151K for $n$ = 6) are far above the FM-transition temperature of $\sim$ 30 K of the bulk YTO, and thus signify that these magnetic screening centers are not directly related to the bulk low temperature ferromagnetism of  YTO.

\subsection{Origin of the $n$-dependent T$_{K}$}

In this part we discuss the reason why T$_K$ can be tuned by the STO layer thickness  (from T$_{K,n = 2}$ $\sim$ 16~K to T$_{K,n = 3}$ $\sim$ 90~K, then to T$_{K,n = 6}$ $\sim$ 151~K, see Fig.~3b in the main text). Considering $\rho(E_{F})$ is a weak decreasing function with STO thickness $n$ (see Fig.~S4(b)) in trilayer  3LTO/$n$STO/3YTO ($n$ = 2, 3, 6), it is seen that $|J|$ is the main control parameter to determine $T_{K}$ with the Eq.~S1. Therefore, it is known naturally that $|J_{n=2}|$ $<$ $|J_{n=3}|$ $<$ $|J_{n=6}|$ with the experimental T$_{K,n=2}$ $<$ T$_{K,n=3}$ $<$ T$_{K,n=6}$. 
On the other hand, due to $|\epsilon_d|$ $\ll$ U (for example, $\epsilon_d$ $\sim$ -0.2~eV for a single YO layer buried in STO and U $\sim$ 4~eV is a physically reasonable value for titanates \cite{Science-2011-Jang}), the expression of $J$ can be simplified as $|J|$ $\propto$ $\frac{|V|^2}{|\epsilon_{d}|}$ with Eq.~S2. Assuming the hybridization $|V|^2$ is a constant \cite{JETP-1995-Irkhin}, it is derived $|J|$ $\propto$ $\frac{1}{|\epsilon_{d}|}$. Therefore, the control parameter to determine T$_{K}$ can be $\epsilon_d$. Thus, $\epsilon_{d,n=2}$ $<$ $\epsilon_{d,n=3}$ $<$ $\epsilon_{d,n=6}$ $<$ $E_{F}$=0. Such a strong ($n$)-dependent behavior of $\epsilon_{d,n=2, 3, 6}$ can be explained well by the weaken octahedra distortion with increasing STO thickness $n$ \cite{PRB-2014-Zhang}. On the other hand, with the competition among the Kondo effect, RKKY and Hund's rule, the Kondo interaction becomes weaker when STO layer thickness is ultra-thin (e.g., 3LTO/2STO/3YTO).

\begin{figure*}[htp]
\includegraphics[width=0.7\textwidth]{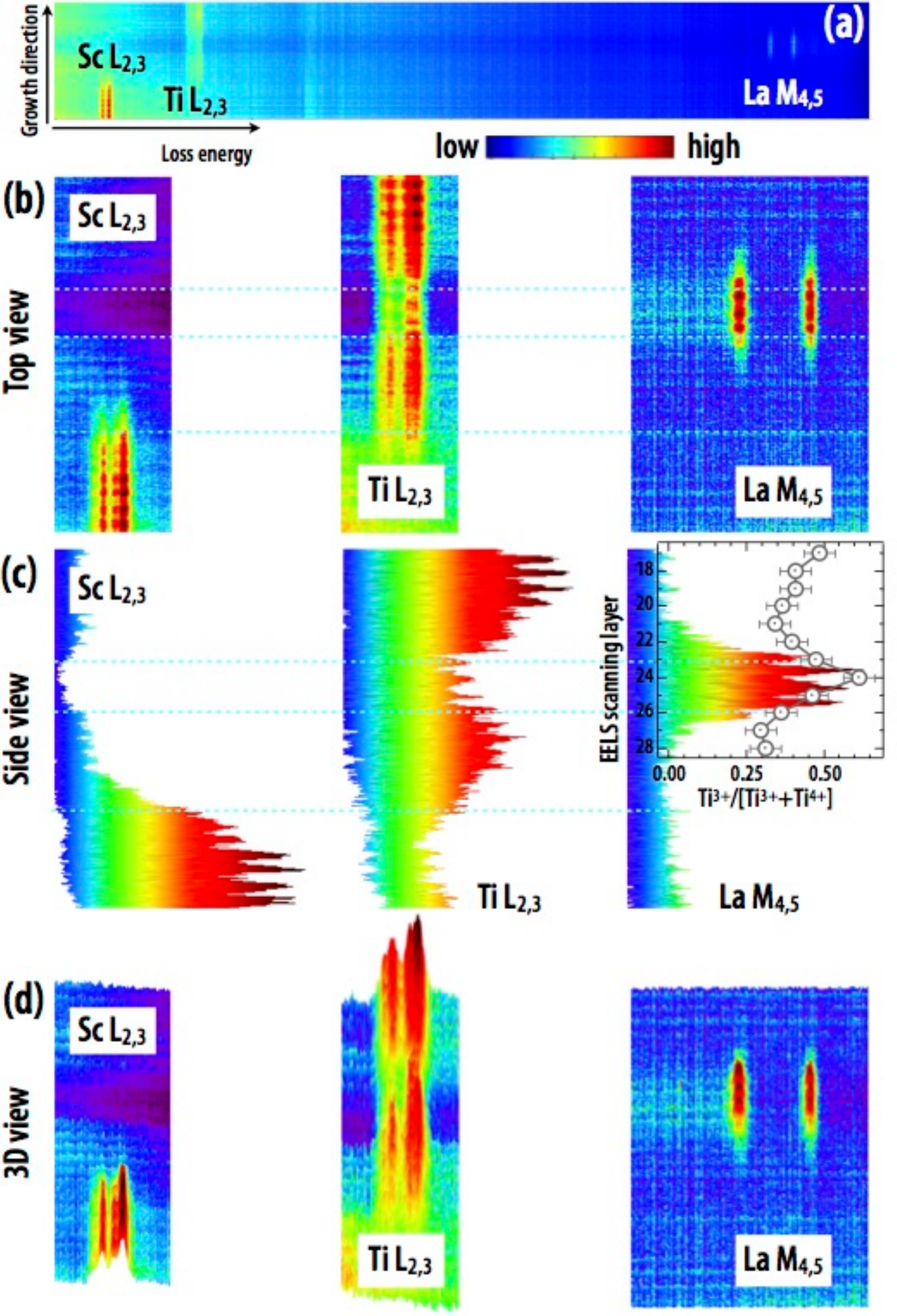}
\caption{\label{} (Color online) \textbf{(Figure S1), EELS mapping.} (a) EELS scanning along the growth direction.  (b) Zoomed EELS spectra at Sc L$_{2,3}$-, Ti L$_{2,3}$-, and La M$_{4,5}$-edges in (a). The dashed cyan lines indicate the sharp interfaces. (c) and (d) Side and 3-dimensional view of (b), respectively. Inset in (c), spatial decay of Ti$^{3+}$ in 3LTO/6STO/3YTO.}
\end{figure*}

\begin{figure*}[htp]
\includegraphics[width=0.8\textwidth]{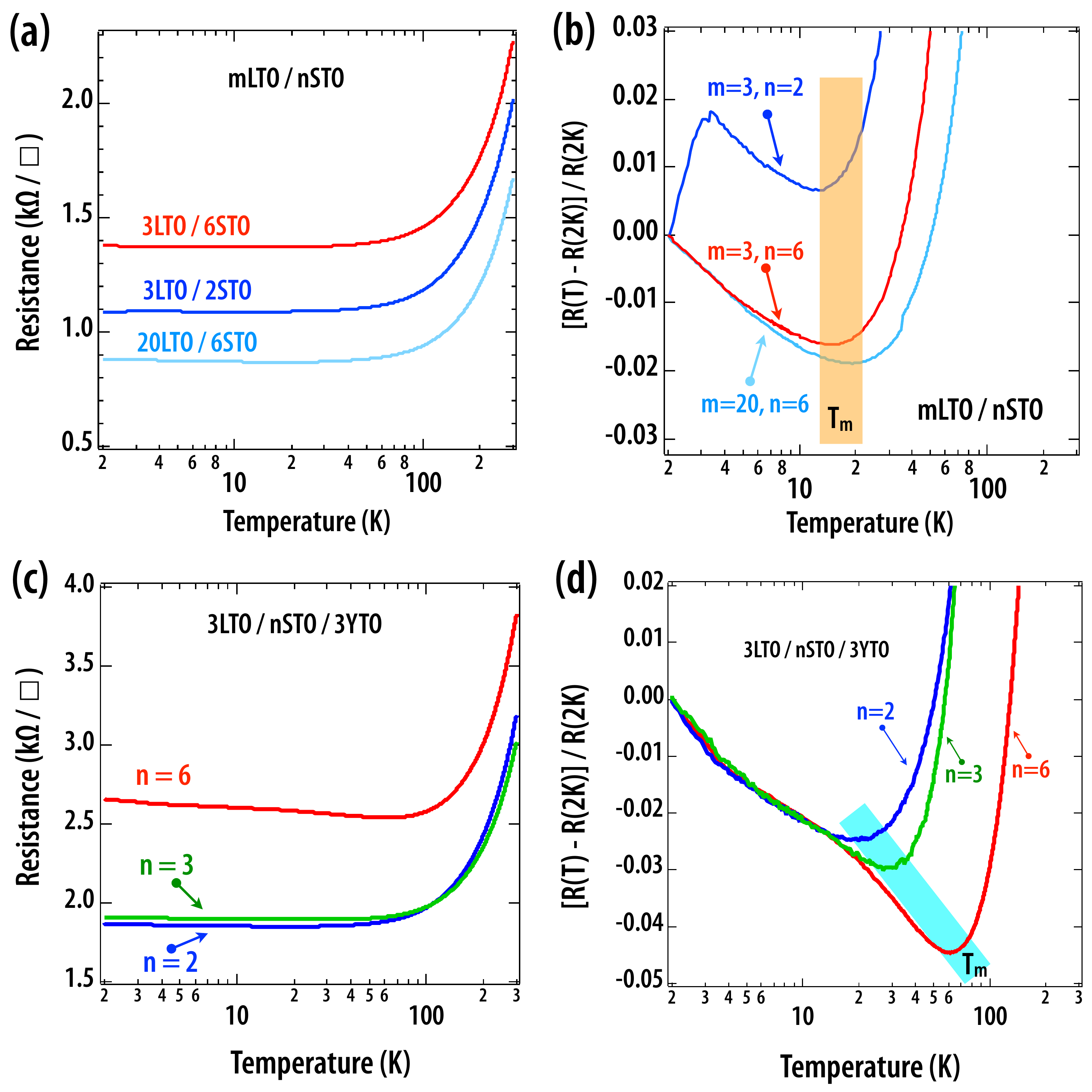}
\caption{\label{} (Color online) \textbf{(Figure S2), Electrical transport.} (a) Electrical sheet resistances (R) of the bicolor SLs mLTO/nSTO  showing a good agreement with the reported values \cite{NC-2010-Bis2}. R~=~R$_{total}$$\times$$I$, where R is the sheet resistance per metallic interface, R$_{total}$ is the sheet resistance of the entire SL, and $I$ is the number of the metallic interfaces. For example, there is 7 metallic LTO/STO interfaces in SL [3LTO/2STO] $\times$ 4, thus $I$~=~7. (b) Normalized sheet resistances of \textbf{a} with the sheet resistances at 2 K, [R(T)-R(2K)]/R(2K). The amplitudes are zoom in $\times$9, $\times$4, and $\times$1 for 3LTO/2STO, 3LTO/6STO, 20LTO/6STO, respectively. The sizes of T$_{m}$ (temperature with the minimum resistance) in bicolor $m$LTO/$n$STO are guided by the yellow shadow behavior, which show a nearly constant behavior. The decreased sheet resistance of 3LTO/2STO below 4~K may be induced by the two-dimensional superconductivity \cite{NC-2010-Bis2}. (c) Sheet resistances (R) of the trilayer SLs 3LTO/$n$STO/3YTO ($n$~=~2,~3,~6). (d) Normalized sheet resistances of (c). The amplitudes of [R(T)-R(2K)]/R(2K) are $\times$2.5, $\times$5, and $\times$1 for $n$~=~2, 3, 6, respectively. The cyan shadow area indicates the values of $T_m$ strongly depend on $n$ (STO layer thickness), which varies as $\sim$~20, $\sim$~28.5, and $\sim$~61~K for $n$~=~2, 3, 6, respectively. In variance to the \textit{bicolor} LTO/STO reference samples with $T_m$ nearly independent of the thickness of individual STO (\textit{n}) or LTO (\textit{m}) layers, the presence of YTO\ in the LTO/$n$STO/YTO SLs gives rise to  a sharp  and systematic increase in $T_m$ with the STO\  layer thickness, \textit{n}.}
\end{figure*}

\begin{figure*}[htp]
\includegraphics[width=0.8\textwidth]{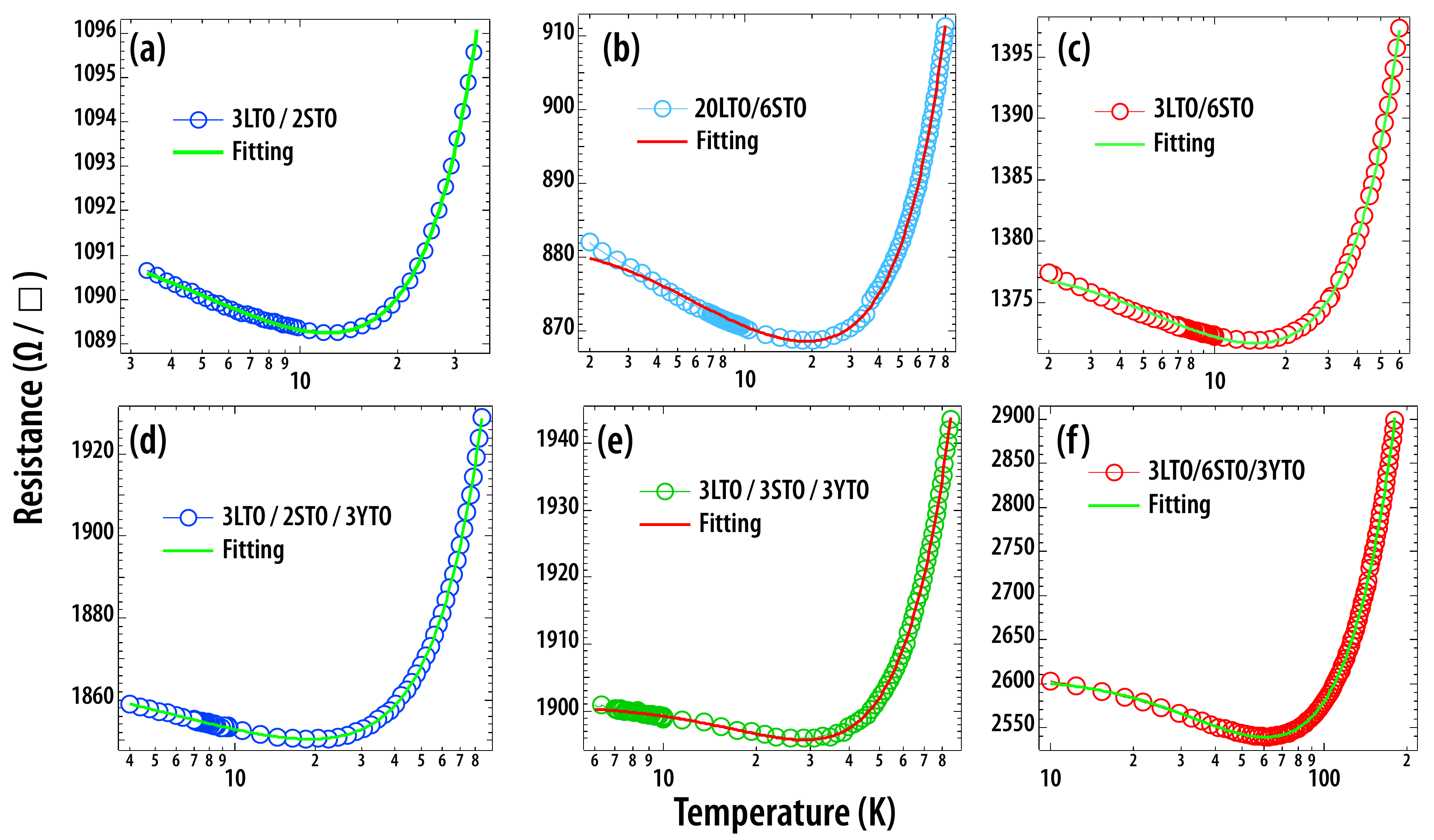}
\caption{\label{} (Color online) \textbf{(Figure S3), Fitting of the sheet resistances with the Kondo effect scenario.} (a)-(f) Comparison of the fitting curves (solid lines, using Eq. S3-S4) with experimental data (circles). (a)-(c), bilayer SLs $m$LTO/$n$STO ($m$~=~3, $n$~=~2; $m$~=20, $n$~=~6 and $m$~=~3, $n$~=~6, respectively). (d)-(f), trilayer SLs 3LTO/$n$STO/3YTO ($n$~=~2, 3, 6). }
\end{figure*}

\newpage
\begin{figure*}[htp]
\includegraphics[width=0.8\textwidth]{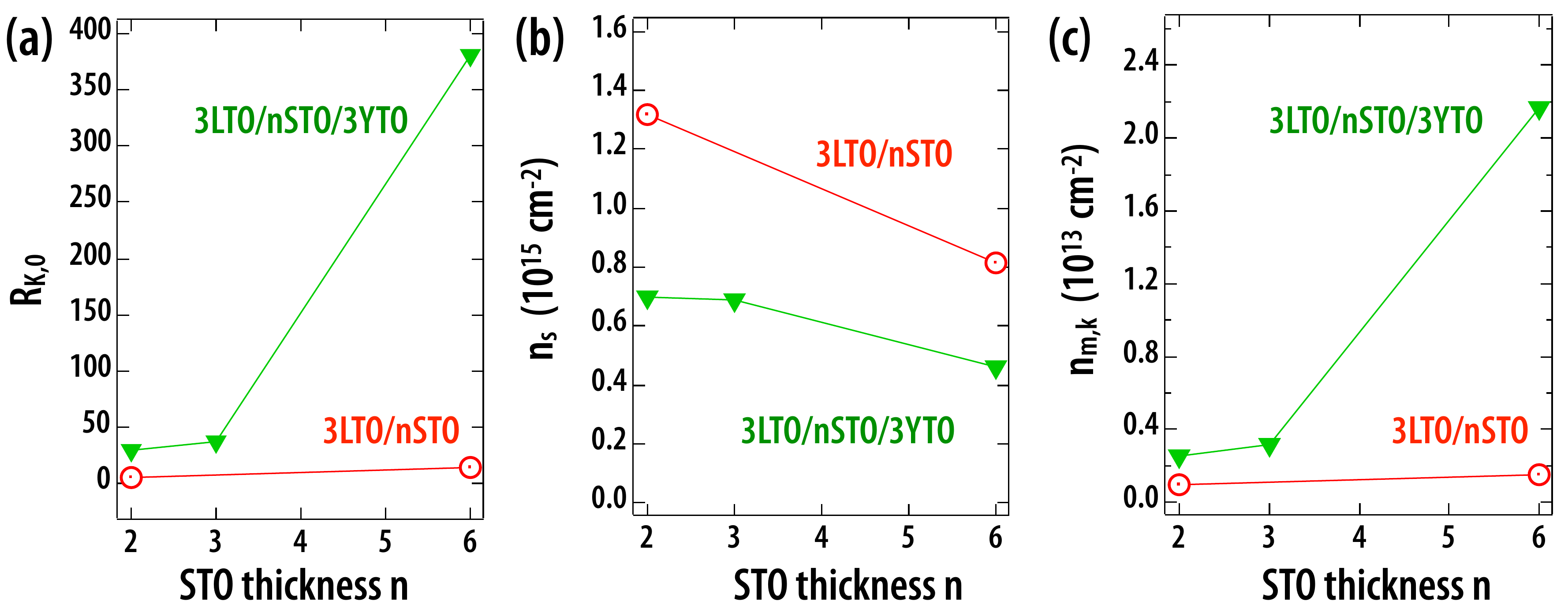}
\caption{\label{} (Color online) \textbf{(Figure S4), Density of conduction electrons and Kondo localized spins.} (a) Kondo resistance near zero temperature extracted from Fig.~S3. (b) Carrier density per interface for bilayer and trilayer SLs measured from Hall resistances at 10~K. (c) Density of the magnetic scattering centers estimated with Eq.~S5.}
\end{figure*}

\begin{figure*}[htp]
\includegraphics[width=0.7\textwidth]{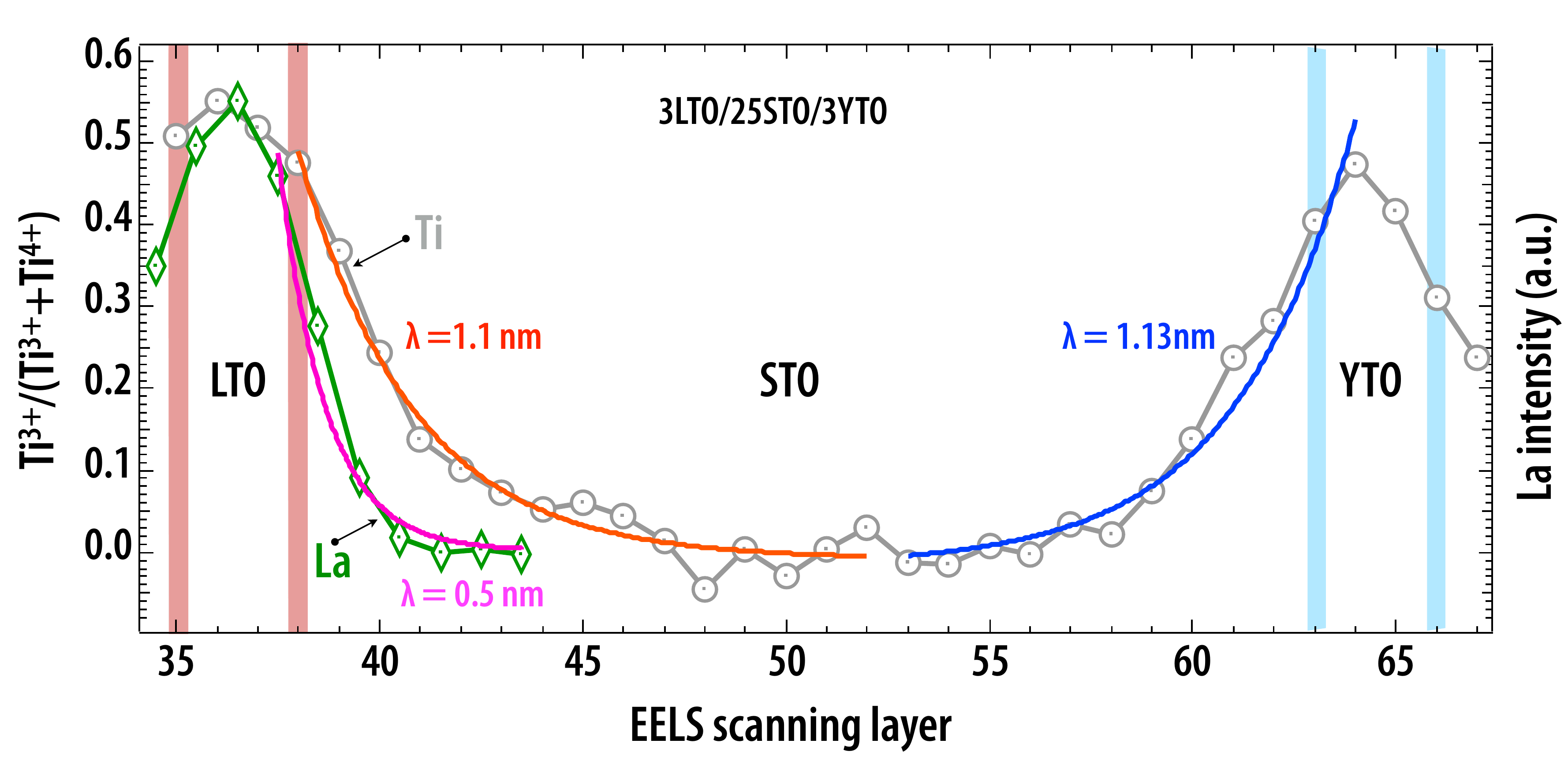}
\caption{\label{} (Color online) \textbf{(Figure S5), Spatial decay of Ti$^{3+}$ and La in 3LTO/25STO/3YTO.} The regular charge-transfer was observed from LTO to STO resulting in 2DEL, while the special \textquotedblleft spin\textquotedblright~transfer was uncovered from YTO to STO leading to the localized spins. The tails of the charge spatial decay can be fitted well by a exponential function decay (solid red and pink lines) $e^{-z/\lambda}$ for the two interfaces, where $z$ is the number of the TiO$_2$-plane or the LaO-plane and $\lambda$ is the electron decay length.}
\end{figure*}

\begin{figure*}[htp]
\includegraphics[width=0.8\textwidth]{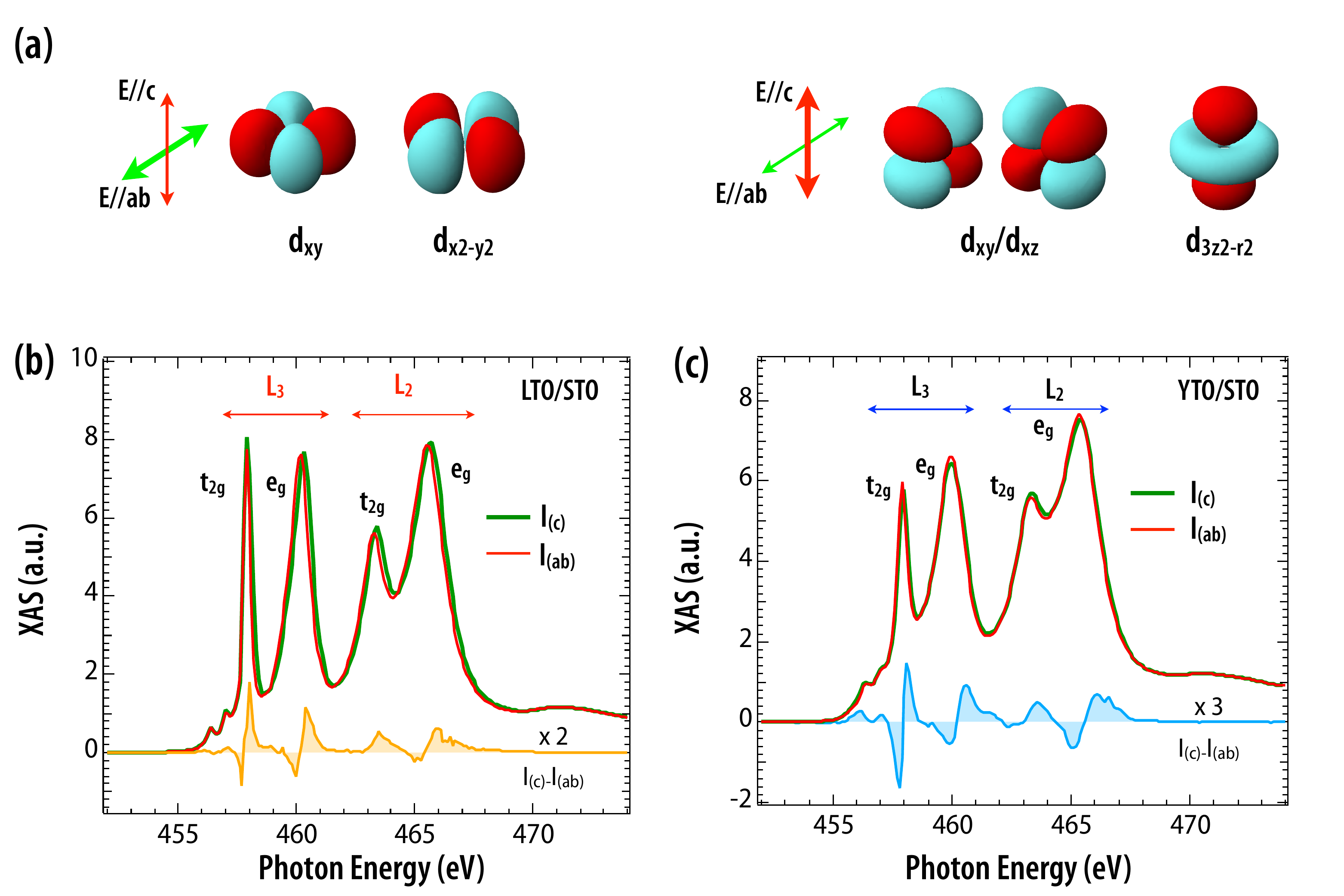}
\caption{\label{} (Color online) \textbf{(Figure S6), Linearly-polarized X-ray absorption spectra.} (a) Schematic of linear polarization vector of the photon $E$ and orbital characters. Out-of-plane [$I(c)$, $E\parallel c$ and $E$ is the linear polarization vector of the photon ] and in-plane [$I(ab)$, $E\parallel ab$] linearly polarized X-ray were used to measure XAS of SLs 3LTO/10STO and 3YTO/2STO at Ti L$_{2,3}$-edge with total electron yield (TEY, surface sensitive) mode at room temperature. The contribution of linearly polarized XAS signal on Ti L$_{2,3}$-edge for t$_{2g}$ (or e$_g$) band arises mainly from the unoccupied Ti $d_{xy}$(or $d_{x^2-y^2}$) [in-plane I$(ab)$] and $d_{xz}$/$d_{yz}$ (or $d_{3z^2-r^2}$)  [out-of-plane I$(c)$] states. It is noted STO is the cap layer for these two samples which guarantees the most contribution of XLD signal is from the STO layer. (b) LTO/STO superlattice. (c) YTO/STO superlattice.}
\end{figure*}

\end{document}